\documentclass[twocolumn,preprintnumbers,amsmath,amssymb]{revtex4}

\usepackage{graphicx}
\usepackage{dcolumn}
\usepackage{bm}

\DeclareMathOperator{\erf}{erf}
\DeclareMathOperator{\erfc}{erfc}
\newcommand{\bra}[1]{\ensuremath{\langle #1 \vert}}
\newcommand{\ket}[1]{\ensuremath{\vert #1  \rangle}}

\renewcommand{\b}[1]{\ensuremath{\mathbf{#1}}}
\newcommand{\lr}{\ensuremath{\text{lr}}}
\newcommand{\sr}{\ensuremath{\text{sr}}}
\newcommand{\unif}{\ensuremath{\text{unif}}}
\newcommand{\md}{\ensuremath{\text{md}}}
\newcommand{\LDA}{\ensuremath{\text{LDA}}}
\newcommand{\Psimu}{\ensuremath{\Psi^{\lr,\mu}}}
\renewcommand{\H}{\ensuremath{\text{H}}}

\newcommand{\rv}{{\bf r}}

\begin{document}

\title{A short-range correlation energy density functional with multi-determinantal reference}

\author{Julien Toulouse}
\author{Paola Gori-Giorgi}
\author{Andreas Savin}
\email{savin@lct.jussieu.fr}
\affiliation{
Laboratoire de Chimie Th\'eorique, CNRS et Universit\'e Pierre et Marie Curie,\\
4 place Jussieu, 75252 Paris, France.
}

\date{\today}

\begin{abstract}
We introduce a short-range correlation density functional defined with respect to a multi-determinantal reference which is meant to be used in a multi-determinantal extension of the Kohn-Sham scheme of density functional theory based on a long-range/short-range decomposition of the Coulomb electron-electron interaction. We construct the local density approximation for this functional and discuss its performance on the He atom.
\end{abstract}

\maketitle

\section{Introduction}
\label{sec:intro}

One of the main difficulties in the Kohn-Sham (KS)~\cite{KohSha-PR-65} scheme of density functional theory (DFT)~\cite{HohKoh-PR-64} is to find approximations for the exchange-correlation energy functional that correctly describe (near-)degeneracy or long-range (e.g., van der Waals) correlation effects. To circumvent this difficulty, a multi-determinantal extension of the KS scheme based on a long-range/short-range decomposition of the Coulomb electron-electron interaction has been proposed~\cite{StoSav-INC-85,SavFla-IJQC-95,Sav-INC-96a,Sav-INC-96,LeiStoWerSav-CPL-97,PolSavLeiSto-JCP-02,SavColPol-IJQC-03,TouColSav-PRA-04,PedJen-JJJ-XX,AngGerSavTou-PRA-XX}. The idea behind this separation is that correlation effects due to the short-range part, involving the correlation cusp, could well be described by the local density approximation (appropriately modified); correlation connected with the long-range part could well be dealt with using standard wave-function methods of quantum chemistry.

In this approach, the ground-state energy of a $N$-electron system in a nuclei-electron potential $v_{ne}(\b{r})$ is obtained in principle exactly by minimization over multi-determinantal wave functions $\Psi$
\begin{eqnarray}
   E &=& \min_{\Psi} \Bigl\{ \bra{\Psi} \hat{T} + \hat{V}_{ne} + \hat{W}_{ee}^{\lr,\mu} \ket{\Psi} 
\nonumber\\
&&+  E_{\H}^{\sr,\mu}[n_{\Psi}] + E_{x}^{\sr,\mu}[n_{\Psi}] + \bar{E}_{c}^{\sr,\mu}[n_{\Psi}] \Bigl\},
\label{Eminmu}
\end{eqnarray}
where $\hat{T}$ is the kinetic energy operator, $\hat{V}_{ne} = \sum_{i} v_{ne}(\b{r}_i)$ is the nuclei-electron interaction operator, $\hat{W}_{ee}^{\lr,\mu} = \sum_{i<j} w_{ee}^{\lr,\mu}(r_{ij})$ is a long-range electron-electron interaction operator with $w_{ee}^{\lr,\mu}(r) = \erf(\mu r)/r$, $E_{\H}^{\sr,\mu}[n] = (1/2) \iint n(\b{r}_1) n(\b{r}_2) w_{ee}^{\sr,\mu}(r_{12}) d\b{r}_1 d\b{r}_2$ is a short-range Hartree functional with $w_{ee}^{\sr,\mu}(r) = \erfc(\mu r)/r$, $E_{x}^{\sr,\mu}[n] = \bra{\Phi[n]} \hat{W}_{ee}^{\sr,\mu} \ket{\Phi[n]} - E_{\H}^{\sr,\mu}[n]$ is a short-range exchange functional where $\hat{W}_{ee}^{\sr,\mu} = \sum_{i<j} w_{ee}^{\sr,\mu}(r_{ij})$ and $\Phi[n]$ is the KS determinant, $\bar{E}_{c}^{\sr,\mu}[n]$ is a short-range correlation functional defined so that Eq.~(\ref{Eminmu}) is exact, and $n_{\Psi}$ is the density coming from $\Psi$. The minimizing wave function in Eq.~(\ref{Eminmu}) will be denoted by $\Psimu$. In these equations, $\mu$ is a parameter controlling the range of the decomposition of the Coulomb interaction. In practice, approximations must be used for the wave function $\Psimu$ and the short-range functionals $E_{x}^{\sr,\mu}[n]$ and $\bar{E}_{c}^{\sr,\mu}[n]$. In particular, the local density approximations have been constructed for $E_{x}^{\sr,\mu}[n]$ and $\bar{E}_{c}^{\sr,\mu}[n]$~\cite{Sav-INC-96,TouSavFla-IJQC-04}. For $\mu=0$, Eq.~(\ref{Eminmu}) reduces to the KS scheme. In fact, in this case, the long-range interaction vanishes, $\hat{W}_{ee}^{\lr,\mu=0}=0$, and the short-range functionals $E_{\H}^{\sr,\mu=0}[n]$, $E_{x}^{\sr,\mu=0}[n]$ and $\bar{E}_{c}^{\sr,\mu=0}[n]$ reduce to the Hartree, exchange and correlation functionals of the KS theory.

Former experience with Eq.~(\ref{Eminmu}) has shown that in general the quality of the wave-function $\Psimu$ obtained with a given approximate functional is much better than that of the functional itself. To extract the maximum information from $\Psimu$, we propose in this work to compute the ground-state energy as
\begin{equation}
E = \bra{\Psimu} \hat{T}  + \hat{V}_{ne} + \hat{W}_{ee} \ket{\Psimu} + \bar{E}_{c,\md}^{\sr,\mu}[n_{\Psimu}],
\label{EPsimu}
\end{equation}
where $\hat{W}_{ee}=\sum_{i<j} 1/r_{ij}$ is the full Coulomb interaction operator and $\bar{E}_{c,\md}^{\sr,\mu}[n]$ is a new short-range correlation functional defined such as Eq.~(\ref{EPsimu}) used with the exact wave function $\Psimu$ is exact. 

We note that Eq.~(\ref{EPsimu}) can formally be made self-consistent by generalizing the ``optimized effective potential'' (OEP) approach (see, e.g., Refs.~\onlinecite{PerEngDreGroGodNogCasMar-BOOK-03,YanAyeWu-PRL-04}) to the multi-determinantal extension of the KS scheme
\begin{equation}
E = \inf_{v} \left\{ \bra{\Psimu[v]} \hat{T}  + \hat{V}_{ne} + \hat{W}_{ee} \ket{\Psimu[v]} + \bar{E}_{c,\md}^{\sr,\mu}[n_{v}] \right\},
\label{Einfvmu}
\end{equation}
where the infinimum is search over one-electron potentials $v(\b{r})$, and $\Psimu[v]$ and $n_{v}$ are, respectively, the ground-state multi-determinantal wave function and density of $\hat{T} + \hat{W}_{ee}^{\lr,\mu} + \sum_i v(\b{r}_i)$. If it exists, the minimizing potential is $v(\b{r})=v_{ne}(\b{r}) + \delta E_{\H}^{\sr,\mu}[n]/\delta n(\b{r}) + \delta E_{x}^{\sr,\mu}[n]/\delta n(\b{r}) + \delta \bar{E}_{c}^{\sr,\mu}[n]/\delta n(\b{r})$, establishing the link with Eq.~(\ref{Eminmu}). Eq.~(\ref{Einfvmu}) reduces to the OEP formulation of the KS theory when $\mu=0$. In this case, in fact, $\Psi^{\lr,\mu=0}$ is the KS determinant and the functional $\bar{E}_{c,\md}^{\sr,\mu=0}[n]$ reduces to the correlation functional of the KS theory. In practice, because the potential corresponding to the functional derivative of $\bar{E}_{c,\md}^{\sr,\mu}[n]$ is small, we expect that the corrections on $\Psimu$ brought by self-consistency are negligible with respect to the errors due to the approximations on $\bar{E}_{c,\md}^{\sr,\mu}[n]$~\cite{SavStoPre-TCA-86}. 



This work is devoted to the study of the functional $\bar{E}_{c,\md}^{\sr,\mu}[n]$: we turn our attention entirely to the correlation energy, without combining it with an approximate functional for exchange. It is thus a pleasure to dedicate this paper to Professor Hermann Stoll who has been a pioneer in the study and application of correlation energy density functionals~\cite{StoPavPre-TCA-78,StoGolPre-TCA-80}. 

The paper is organized as follows. In Sec~\ref{sec:srcmd}, we discuss the short-range correlation functional $\bar{E}_{c,\md}^{\sr,\mu}[n]$ and its relation to the functional $\bar{E}_{c}^{\sr,\mu}[n]$. In Sec.~\ref{sec:lda}, we construct a local density approximation for $\bar{E}_{c,\md}^{\sr,\mu}[n]$. In Sec.~\ref{sec:results}, we assess the accuracy of this approximation for the He atom. Sec.~\ref{sec:conclusion} contains our conclusions.
Atomic units (a.u.) are used throughout this work.

\section{The short-range correlation functional $\bar{E}_{c,\md}^{\sr,\mu}[n]$}
\label{sec:srcmd}

The short-range correlation functional $\bar{E}_{c,\md}^{\sr,\mu}[n]$ in Eq.~(\ref{EPsimu}) is defined with respect to the multi-determinantal wave function $\Psimu$, in contrast to the short-range correlation functional $\bar{E}_{c}^{\sr,\mu}[n]$ in Eq.~(\ref{Eminmu}) defined with respect to the one-determinant wave function $\Phi$. It is easy to see that these two functionals are related to each other through
\begin{eqnarray}
\bar{E}_{c,\md}^{\sr,\mu}[n] = \bar{E}_c^{\sr,\mu}[n] + \Delta^{\lr-\sr,\mu}[n],
\label{Ecsrres}
\end{eqnarray}
where
\begin{eqnarray}
\Delta^{\lr-\sr,\mu}[n] =
\nonumber\\
 - \left( \bra{\Psimu[n]} \hat{W}_{ee}^{\sr,\mu} \ket{\Psimu[n]} - \bra{\Phi[n]} \hat{W}_{ee}^{\sr,\mu} \ket{\Phi[n]} \right).
\label{DEcsrres}
\end{eqnarray}
The quantity $\Delta^{\lr-\sr,\mu}$ vanishes for $\mu=0$ and $\mu \to \infty$.

It is interesting to study the behavior of $\bar{E}_{c,\md}^{\sr,\mu}[n]$ in the limit of a very short-range interaction, i.e. when $\mu \to \infty$.
In this limit, the short-range interaction behaves as~\cite{TouColSav-PRA-04}
\begin{eqnarray}
w_{ee}^{\sr,\mu}(r) =  \frac{\pi}{\mu^2} \delta(\b{r}) + O\left(\frac{1}{\mu^3}\right),
\end{eqnarray}
leading to the following asymptotic expansion of $\bar{E}_c^{\sr,\mu}$~\cite{PolColLeiStoWerSav-IJQC-03,SavColPol-IJQC-03,TouColSav-PRA-04}
\begin{eqnarray}
\bar{E}_c^{\sr,\mu\to\infty} = \frac{\pi}{2\mu^2} \int n_{2,c}(\b{r},\b{r}) d\b{r} + O\left(\frac{1}{\mu^3}\right),
\label{Ecsrmuinf}
\end{eqnarray}
where $n_{2,c}(\b{r},\b{r})$ is the correlation on-top pair density. The asymptotic expansion of $\Delta^{\lr-\sr,\mu}$ as $\mu\to\infty$ is obtained similarly from its definition, Eq.~(\ref{DEcsrres}), leading to
\begin{eqnarray}
\Delta^{\lr-\sr,\mu\to\infty} = -\frac{\pi}{2\mu^2} \int n_{2,c}(\b{r},\b{r}) d\b{r} + O\left(\frac{1}{\mu^3}\right).
\label{Dlrsrmuinf}
\end{eqnarray}
The first terms in Eqs.~(\ref{Ecsrmuinf}) and~(\ref{Dlrsrmuinf}) cancel, and therefore $\bar{E}_{c,\md}^{\sr,\mu}$ decays at least as $1/\mu^3$ when $\mu\to\infty$.



\section{Local density approximation}
\label{sec:lda}

A local density approximation (LDA) can be constructed for $\bar{E}_{c,\md}^{\sr,\mu}[n]$ 
\begin{eqnarray}
\bar{E}_{c,\md,\LDA}^{\sr,\mu}[n] = \int n(\b{r}) \bar{\varepsilon}_{c,\md,\unif}^{\sr,\mu}(n(\b{r})) d\b{r},
\label{eq_mdLDA}
\end{eqnarray}
where the corresponding correlation energy per particle in the uniform electron gas $\bar{\varepsilon}_{c,\md,\unif}^{\sr,\mu}(n)$ is given by
\begin{equation}
\bar{\varepsilon}_{c,\md,\unif}^{\sr,\mu}(n)   =  \bar{\varepsilon}_{c,\unif}^{\sr,\mu}(n) 
+\Delta^{\lr-\sr,\mu}_{\unif}(n).
\label{eq_defLDAecmd}
\end{equation}
In Eq.~(\ref{eq_defLDAecmd}), $\bar{\varepsilon}_{c,\unif}^{\sr,\mu}(n)$ is the correlation energy per particle defining the LDA approximation for $\bar{E}_{c}^{\sr,\mu}[n]$ (see Refs.~\onlinecite{Sav-INC-96,TouSavFla-IJQC-04}), and $\Delta^{\lr-\sr,\mu}_{\unif}(n)$ is given by
\begin{equation}
\Delta^{\lr-\sr,\mu}_{\unif}(n)=- \frac{n}{2} \int_{0}^{\infty} g_{c,\unif}^{\lr,\mu}(r,n) w_{ee}^{\sr,\mu}(r) 4 \pi r^2 dr,
\label{ecunifres}
\end{equation}
where $g_{c,\unif}^{\lr,\mu}(r,n)$ is the correlation pair-distribution function of a uniform electron gas with long-range interaction $w_{ee}^{\lr,\mu}(r)$ and density $n$. The correlation hole of this ``long-range'' electron gas is then given by $n\,g_{c,\unif}^{\lr,\mu}(r,n)$.

Since an estimate of the energy $\bar{\varepsilon}_{c,\unif}^{\sr,\mu}(n)$ from coupled-cluster calculations is available~\cite{Sav-INC-96,TouSavFla-IJQC-04}, we only need to compute the term $\Delta^{\lr-\sr,\mu}_{\unif}(n)$ to build the LDA functional of Eq.~(\ref{eq_mdLDA}). In order to estimate $\Delta^{\lr-\sr,\mu}_{\unif}(n)$  we proceed as follows. We first notice that $w_{ee}^{\sr,\mu}(r)=\erfc(\mu r)/r$ in the integrand of Eq.~(\ref{ecunifres}) only samples the part of $g_{c,\unif}^{\lr,\mu}(r,n)$ corresponding to $r\lesssim 1/\mu$. For the standard uniform electron gas (with full interaction $1/r$) the ``extended Overhauser model''~\cite{GorPer-PRB-01} proved to be able to yield accurate results for  $g_{c,\unif}(r,n)$ in the short-range region defined by  $r\le r_s$, where $r_s=(4 \pi\,n/3)^{-1/3}$. We can thus use this simple model to calculate $g_{c,\unif}^{\lr,\mu}(r,n)$  and to produce an estimate for $\Delta^{\lr-\sr,\mu}_{\unif}(n)$ that should be reliable for $\mu$-values for which $\mu r_s\gtrsim 1$.

The scattering equations of the ``extended Overhauser model'' are widely explained in Refs.~\onlinecite{GorPer-PRB-01,DavPolAsgTos-PRB-02}. Here we solved the same equations with the electron-electron interaction $\erf(\mu r)/r$ screened by a sphere of radius $r_s$ of uniform positive charge density $n$ and attracting the electrons with the same modified interaction,
\begin{equation}
V_{\rm eff}(r,r_s,\mu)=\frac{\erf(\mu r)}{r}
-\int_{|\rv'|\le r_s} n\,  \frac{\erf(\mu|\rv' - \rv|)}{|\rv' - \rv|}\,d \rv'.
\label{eq_veff}
\end{equation}
This potential is reported in the Appendix of Ref.~\onlinecite{GorSav-PRA-05}, where it has been used for two-electron atoms with very accurate results for the corresponding short-range correlation energy. $V_{\rm eff}(r,r_s,\mu)$ is a screened potential that tends to the ``Overhauser potential''~\cite{Ove-CJP-95,GorPer-PRB-01} when $\mu\to\infty$, and which goes to zero when $\mu\to 0$. As in the original work of Overhauser~\cite{Ove-CJP-95}, the idea behind Eq.~(\ref{eq_veff}) is that the radius of the screening ``hole'' is exactly equal to $r_s$.

\begin{figure}
\includegraphics[scale=0.7]{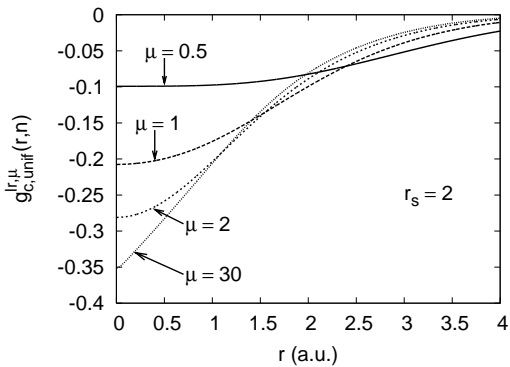}
\caption{A sample of the pair-correlation functions $g_{c,\unif}^{\lr,\mu}(r,n)$ of a uniform electron gas of density $3/(4\pi r_s^3)$ (here $r_s=2$) with long-range-only interaction $\erf(\mu r)/r$ obtained from the ``extended Overhauser model''~\cite{GorPer-PRB-01} [see Eq.~(\ref{eq_veff})]. The correlation hole is given by $n\,g_{c,\unif}^{\lr,\mu}(r,n)$.}
\label{fig_gc}
\end{figure}

A sample of the pair-correlation functions $g_{c,\unif}^{\lr,\mu}(r,n)$ that we have obtained is reported in Fig.~\ref{fig_gc}. As long as $\mu$ is not large we clearly see the absence of the cusp [$g_{c,\unif}'(r=0)$ is not zero  for a system with interaction $1/r$ at small $r$, but it is zero for the $\erf(\mu r)/r$ interaction]. As expected, as $\mu$ increases the hole deepens, and for very large $\mu$ we see that the cusp starts to appear.

\begin{table*}
\begin{tabular}{lll|lll|lll|lll|lll}
\hline\hline
$\mu$ & $r_s$ & $\;\;\Delta^{\lr-\sr,\mu}_{\unif}$ & $\mu$ & $r_s$ & $\;\;\;\Delta^{\lr-\sr,\mu}_{\unif}$ & $\mu$ & $r_s$ & $\;\;\;\Delta^{\lr-\sr,\mu}_{\unif}$ & $\mu$ & $r_s$ & $\;\;\;\Delta^{\lr-\sr,\mu}_{\unif}$  & $\mu$ & $r_s$ & $\;\;\;\Delta^{\lr-\sr,\mu}_{\unif}$\\
\hline
 2 &  0.5 & $0.0173$  &     3 &  0.5 & $0.0136$  &  5 & 0.2 & $0.0175$   & 10 &  0.2 & $0.0109$                       &   15  & 0.2 & $0.00683$            \\
 2 &  1 & $ 0.00979$ &     3 &  1 & $0.00585$    & 5 &  0.5 & $0.00813$  & 10 &  0.5 & $0.00299$      & 15  & 0.5 & $0.00151$            \\
2 &  2 & $ 0.00310$ &     3 &  2 & $0.00153$   & 5 &  1 & $0.00265$  & 10 &  1 & $0.000776$          & 15 &  1 & $0.000363$             \\
2 &  3 &  $0.00125$ &     3 &  3 & $0.000586$  & 5 & 2 & $0.000598$   & 10 &  2 & $0.000158$               & 15 & 2 & $7.16\cdot 10^{-5}$        \\
2 &  4 & $0.000608$ &     3 &  4 & $0.000278$  & 5 &  3 & $0.000219$   & 10 &  3 & $5.63\cdot 10^{-5}$   & 15 &  3 & $2.52\cdot 10^{-5}$           \\
2 &  5 & $ 0.000335$ &    3 &  5 & $0.000151$  & 5 &  4 & $0.000102$     & 10 &  4 & $2.59\cdot 10^{-5}$   &  15 &  4 & $1.16\cdot 10^{-5}$            \\
2 &  6 & $0.000202$  &  3 &  6 & $9.08\cdot 10^{-5}$  &  5 &  5 & $ 5.52\cdot 10^{-5}$ & 10 &  5 & $1.39\cdot 10^{-5}$   & 15 &  5 & $6.19\cdot 10^{-6}$            \\
\hline\hline
\end{tabular}
\caption{A sample of the values of $\Delta^{\lr-\sr,\mu}_{\unif}$ [see Eqs.~(\ref{eq_defLDAecmd})-(\ref{ecunifres})] computed from the extended Overhauser model~\cite{GorPer-PRB-01}.}
\label{tab_delta}
\end{table*}

Some of the values of $\Delta^{\lr-\sr,\mu}_{\unif}(n)$ for $\mu r_s\gtrsim 1$ are reported in Table~\ref{tab_delta}. An estimate of $\Delta^{\lr-\sr,\mu}_{\unif}(n)$ in the region  not accessible with the extended Overhauser model, $\mu r_s\lesssim 1$, has been obtained by a simple interpolation between our data and zero, since, as explained in Sec.~\ref{sec:srcmd}, $\Delta^{\lr-\sr,\mu}$ vanishes when $\mu\to 0$. 
In the opposite limit, $\mu \to\infty$, $\Delta^{\lr-\sr,\mu}_{\unif}(n)$ behaves as in Eq.~(\ref{Dlrsrmuinf}), which for a system of uniform density reads 
\begin{equation}
\Delta^{\lr-\sr,\mu\to\infty}_{\unif}(n)=-\frac{3\, g_{c,\unif}(0,n)}{8\, r_s^3\,\mu^2}+O\left(\frac{1}{\mu^3}\right),
\end{equation}
where $g_{c,\unif}(0,n)$ is the on-top value ($r=0$) of the pair-correlation function of the Coulombic uniform electron gas of density $n$~\cite{GorPer-PRB-01}. We found that the $\Delta^{\lr-\sr,\mu}_{\unif}(n)$ computed with the extended Overhauser model accurately recover this limiting behavior.

For future applications, a more accurate LDA functional for $\Delta^{\lr-\sr,\mu}$ (especially for $\mu r_s \lesssim 1$) will be available from quantum Monte Carlo calculations~\cite{PazMorGorBac-xxx-xx}.

\section{Results for the He atom}
\label{sec:results}

For the He atom, the short-range correlation energies $\bar{E}_c^{\sr,\mu}$ and $\bar{E}_{c,\md}^{\sr,\mu}$ have been calculated with a precision of the order of $1$ mH as follows. An accurate density is calculated at the full configuration interaction level with a large Gaussian basis set and the optimization of the potential in the Legendre transform formulation~\cite{Lie-IJQC-83,NalPar-JCP-82} of density functionals enables to compute accurately the correlation energy $\bar{E}_c^{\sr,\mu}$ associated to that density (see Refs.~\onlinecite{ColSav-JCP-99,PolColLeiStoWerSav-IJQC-03,TouColSav-PRA-04} for details). The corresponding accurate multi-determinantal wave function $\Psimu$ and the KS wave function $\Phi$ are also obtained in this procedure, which give access to an accurate evaluation of  $\Delta^{\lr-\sr,\mu}$ and consequently of $\bar{E}_{c,\md}^{\sr,\mu}$.

\begin{figure}
\includegraphics[scale=0.7]{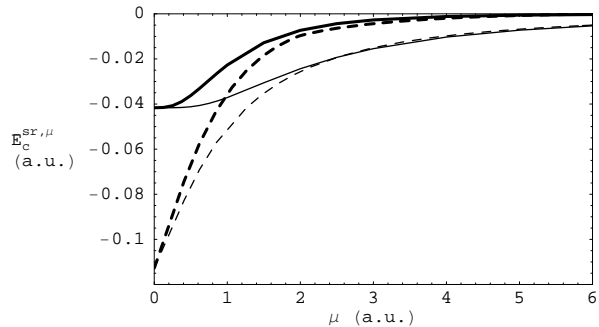}
\caption{Short-range correlation energies $\bar{E}_c^{\sr,\mu}$ (thin curves) and $\bar{E}_{c,\md}^{\sr,\mu}$ (thick curves) with respect to $\mu$ for the He atom. Accurate calculations (solid curves) are compared to the LDA approximation (dashed curves).
}
\label{fig:EcsrmuEcsrmures-he-erf}
\end{figure}

\begin{table*}
\begin{tabular}{c | r r r r r r r r r}

\hline\hline
$\mu$  & 0.00 & 0.25 & 0.50 & 0.75 & 1.00 & 1.50 & 2.00 & 3.00 & 5.00\\
\hline
$\Delta \bar{E}_c^{\sr,\mu}$ & -0.071 & -0.053 & -0.035 & -0.022 & -0.014 & -0.005 & -0.001 & 0.000 & 0.000 \\
$\Delta \bar{E}_{c,\md}^{\sr,\mu}$ & -0.071 & -0.048 & -0.030 & -0.019 & -0.013 & -0.006 & -0.002 & -0.002 & 0.000 \\
\hline\hline
\end{tabular}
\caption{LDA errors $\Delta \bar{E}_c^{\sr,\mu} = \bar{E}_{c,\LDA}^{\sr,\mu} - \bar{E}_c^{\sr,\mu}$ and $\Delta \bar{E}_{c,\md}^{\sr,\mu} = \bar{E}_{c,\md,\LDA}^{\sr,\mu} - \bar{E}_{c,\md}^{\sr,\mu}$ with respect to $\mu$ for the He atom.
}
\label{tab:DEcsrmuEcsrmures-he-erf}
\end{table*}

The accurate and LDA short-range correlation energies $\bar{E}_{c}^{\sr,\mu}$ and $\bar{E}_{c,\md}^{\sr,\mu}$ are compared in Fig.~\ref{fig:EcsrmuEcsrmures-he-erf}. For all values of $\mu$, we have $|\bar{E}_{c,\md}^{\sr,\mu} |< |\bar{E}_{c}^{\sr,\mu}|$, meaning that in Eq.~(\ref{DEcsrres}) $\bra{\Psimu} \hat{W}_{ee}^{\sr,\mu} \ket{\Psimu} < \bra{\Phi} \hat{W}_{ee}^{\sr,\mu} \ket{\Phi}$ which seems natural for a repulsive interaction. Tab.~\ref{tab:DEcsrmuEcsrmures-he-erf} compare the LDA errors on $\bar{E}_{c}^{\sr,\mu}$ and $\bar{E}_{c,\md}^{\sr,\mu}$. One sees that the LDA errors for this two short-range correlation energies are of the same order of magnitude for all values of $\mu$.

\section{Conclusions}
\label{sec:conclusion}

In this work, we have reexamined the multi-determinantal extension of the KS scheme based on a long-range/short-range decomposition of the Coulomb electron-electron interaction. Contrary to previous works where the short-range correlation functional was defined with respect to the KS determinant, we have introduced a new short-range correlation functional defined with respect to the multi-determinantal wave function. We have constructed the local density approximation for this new functional. The example of the He atom suggests that the local density approximation is essentially as accurate as for the short-range correlation functional defined with respect to the KS determinant. We believe that this work paves the way to a multi-determinantal extension of the KS scheme using a correlation-only density functional.


\bibliographystyle{apsrev}
\bibliography{biblio}

\end{document}